# Realizing One-dimensional Metallic States in Graphene via Periodically Coupled Zeroth Pseudo-Landau Levels

Yi-Wen Liu[1,§], Zhen Zhan[2,§], Zewen Wu[2], Chao Yan[1], Shengjun Yuan[2,†], and Lin He[1,†]

[1]Center for Advanced Quantum Studies, Department of Physics, Beijing Normal University, Beijing, 100875, People's Republic of China

[2]Key Laboratory of Artificial Micro- and Nano-structures of the Ministry of Education and School of Physics and Technology, Wuhan University, Wuhan 430072, China

[§]These authors contributed equally to this work.

[†]Correspondence and requests for materials should be addressed to Shengjun Yuan (email: s.yuan@whu.edu.cn) and Lin He (e-mail: helin@bnu.edu.cn).

**Strain-induced pseudo-magnetic fields can mimic real magnetic fields to generate a zero-magnetic-field analogue of the Landau levels (LLs), *i.e.*, the pseudo-LLs, in graphene. The distinct nature of the pseudo-LLs enables one to realize novel electronic states beyond that can be feasible with real LLs. Here, we report the realization of one-dimensional (1D) metallic states, which can be described well by the Su-Schrieffer-Heeger model, in graphene via periodically coupled zeroth pseudo-LLs. In our experiment, nanoscale strained structures embedded with pseudo-LLs are generated periodically along 1D channel of suspended graphene monolayer. Our experiments demonstrate that the zeroth pseudo-LLs of these strained structures are coupled to form metallic states, exhibiting a serpentine pattern that snakes back and forth along the 1D suspended graphene monolayer. These results are verified theoretically by large-scale tight-binding calculations of the strained samples. Our result provides a new pathway to realize novel quantum states and engineer the electronic properties of graphene by using the localized pseudo-LLs as building blocks.**

A non-uniform strain in graphene can create pseudo-magnetic fields (PMFs), which generate a unique Landau quantization at zero external magnetic field [1-6]. By tuning deformed structure of graphene, the PMFs can be changed from zero to hundreds of Tesla, providing an efficient way to tailor electronic properties of graphene [7-24]. Therefore, throughout the last ten years many efforts have been made to introduce custom-designed strained structures [25-28] and PMFs in graphene [7-24]. Even more interestingly, the different nature between the PMFs and the real magnetic fields enables one to realize novel quantum states that cannot be realized purely with the real magnetic fields [14-19,23]. For example, a special valley-polarized Landau quantization was observed in strained graphene because the PMFs point in opposite directions at the two different valleys of graphene [14,19]. The strain-induced pseudo-Landau levels (pseudo-LLs), which are also quite different from real LLs, only depend on the deformed structure of graphene and, thus, can be localized at nanoscale. In this Letter, by taking advantage of the nanoscale localization of the pseudo-LLs, we demonstrated for the first time that it is possible to use the pseudo-LLs as the building block to realize a model system exhibiting novel quantum states. We realized one-dimensional (1D) metallic states, which can be described well by the Su-Schrieffer-Heeger (SSH) model [29,30], via periodically coupled zeroth pseudo-LLs in graphene. These robust 1D SSH metallic states in graphene were directly visualized in our scanning tunneling microscopy (STM) measurements and justified by a full tight-binding calculation.

In our experiment, nanoscale strained structures are generated periodically along 1D channel of suspended graphene monolayer. To obtain the 1D channel of suspended graphene monolayer, we adopted an etching technique of graphene using high-flow of hydrogen and metallic nanoparticles [31-37]. According to our STM measurements, the width of the 1D channels, which is determined by the diameter of the metallic nanoparticles, ranges from a few nanometers to several tens nanometers (Fig. S1). Two types of the 1D channels are obtained after the etching processes, as summarized in Fig. 1. The first one is that the topmost graphene is etched and 1D nanoscale trench with atomic-scale sharp edges of the topmost graphene is obtained, as shown in Figs. 1(a) and 1(c). The second type of the 1D channels is that the underlying graphene layers are etched, leaving the topmost graphene layer untouched, as shown in Figs. 1(b) and 1(e). The two types of the 1D channels are quite different in the heights of the trenches and the strengths of the intervalley scattering around them, as shown in Figs. 1(d), 1(f) and Fig. S1. Therefore, it is quite easy to distinct them in the experiment. In this work, we mainly focused on the second type of the channels because that it provides 1D nanoscale suspended graphene monolayer, which is vital to realize 1D periodic strained structures, as illustrated subsequently.

For most of the 1D suspended graphene membrane (the 1D structure above large Ni terraces or multilayer graphene terraces), the length is over several hundreds of nanometers and there is no observable strained structure (Fig. S1), indicating that the supporting substrate does not introduce stress on the 1D suspended structure even though there are different thermal expansion coefficients between the topmost graphene monolayer and the substrate. However, for the 1D suspended graphene monolayer above terraces with small widths (see Figs. S2 and S3 for more experimental characterizations), we observed clear strained structures along the 1D structure, as shown in Figs. 2(a)-2(c). In this case, the 1D suspended graphene monolayer is suddenly cut off at the edges of the terraces and, therefore, the length of the 1D structure is relatively short. The two ends at the edges of the terraces may introduce strain in the short 1D suspended graphene monolayer and, consequently, generate periodic strained structures along it (see Fig. S4 for more experimental data), as shown in Fig. 2(b).

In strained graphene, lattice deformation can effectively modulate electron hopping and create pseudo-LLs around Dirac point of graphene. The strain-induced pseudo-LLs can be explicitly demonstrated by scanning tunneling spectroscopy (STS) measurements in which the pseudo-LLs are clearly observed as pronounced peaks [7-24]. Figure 2(d) shows representative STS spectra acquired on different positions of a strained structure in the 1D suspended graphene monolayer (marked in Fig. 2(c)). The four pronounced peaks in the spectra are attributed to the strain-induced pseudo-LLs, which can be fitted well by the Landau quantization of massless Dirac fermions in graphene monolayer (see Fig. S3). In our experiment, a robust zeroth pseudo-LL, exhibiting a large full width about 0.2 eV, is observed in all the spectra of the strained graphene and the signals of high pseudo-LLs with LL index $n \neq 0$ are much weaker. Theoretically, the description of strained graphene as an effective PMF is exactly valid only at the charge neutrality point and fails at higher energies [14]. Therefore, the high pseudo-LLs are not well defined and, consequently, much weaker in the spectra. To measure the spatial distribution of the PMFs, STS spectra at different positions of the 1D suspended graphene monolayer are carefully measured. The values of the PMFs in each position of the strained structures can be obtained according to the fitting of the Landau quantization of massless Dirac fermions. Figure 2(e) shows distribution of the PMFs measured around four strained ripples along the 1D suspended graphene monolayer. Obviously, the PMFs are quite non-uniform and show similar period as the strained structures along the 1D structure.

To further explore effects of the quasi-periodic distributed PMFs on electronic properties of the 1D suspended graphene monolayer, conductance maps, which reflect spatial-distributed local density of states (LDOS), at different energies are measured, as shown in Figs. 3(a)-(c) (see Fig. S5 for maps at other energies). At energies, both positive and negative, far away from the zeroth pseudo-LL, there is almost no observable feature induced by the strained structures in the 1D suspended graphene monolayer (Fig. 3 (c)). By reducing the energy to close to the zeroth pseudo-LL, strain-induced electronic states, which are mainly localized in the strained structures, are clearly observed along the 1D structure (Fig. 3(a)). At the zeroth pseudo-LL, we observed a serpentine pattern that snakes back and forth along the 1D suspended graphene monolayer, which reminds us the 1D SSH metallic states observed in graphene nanoribbons very recently [38-40]. Such a result indicates that the zeroth pseudo-LLs at different strained structures are coupled to form a 1D metallic state.

The serpentine zeroth pseudo-LLs in the 1D suspended graphene are further verified in theory by utilizing a large-scale propagation method in the frame of tight-binding model. The strained graphene used in the calculation is composed of transverse ($x$) buckling and longitudinal (*y*) ripples (see SI Note 9 and Fig. S6 for details). The out-of-plane displacement of the strained structure along the 1D suspended graphene monolayer is modelled by a sinusoidal function. We vary five setting parameters of buckling and ripples to mimic the strained structure in Fig. 2(b), and obtain the structure 1 in Fig. 4(a) (SI Note 9). The distortion-induced strain is defined as $\varepsilon = (\frac{\sum_i d_i}{3d} - 1) \times 100\%$ with $d_i$ (*i* =1,2,3) the first neighbor interatomic distance in the deformed lattice and *d* = 0.142 nm the distance between carbon atoms in equilibrium. As illustrated in Fig. 4(b), the ε shows a periodicity and with the maximum value around 4% located around the crossover (ripple head) of the buckling and ripples. The strength of the strain-induced PMFs in the 1D suspended graphene monolayer are evaluated in Fig. 4(c). The PMFs vary at different positions and exhibit the same period as the strained structure, which is in good agreement with the experimental results. The strain-induced PMFs also strongly modify electronic properties of the 1D suspended graphene monolayer. Figure 4(d) shows representative LDOS at two different positions of the strained structure. The strain-induced pseudo-LLs can be explicitly identified, as observed in our experiment. To further compare with the experimental result, we calculated energy-fixed LDOS of the 1D suspended graphene monolayer, as shown in Figs. 3(d)-(f). Similar to the experimental results, the 1D metallic states, exhibiting a serpentine pattern, are only observed at the zeroth pseudo-LL energy.

In the strained structures of the 1D suspended graphene layer, the maximum value of strain exists near the ripple head and the maximum response of the LDOS is around the highest point of the ripple. Consequently, this strain induced serpentine pattern can be understood by a low-energy effective SSH tight-binding model [29,30]. The dispersion relationship of the two-sublattice system in the SSH model is described as $E_\pm(k) = \pm\sqrt{|t_1|^2 + |t_2|^2 + 2|t_1||t_2|cos(k + \delta)}$, where $\delta$ is the relative phase between two effective hopping terms $t_1$ and $t_2$. When $t_1 = t_2 = t$, a metallic state forms in the chain. In our experiment, the staggered strained structures mimic different sublattices and the robust pseudo-LLs contribute to the localized zero modes. The overlap of the adjacent zero-mode states with equal hopping parameters $t_1 = t_2 \approx 50$ meV leads to an enlarged SSH metallic chain in the channel in Fig. 3 (here the bottom of zero pseudo-LLs are defined as the effective band width in Figs. 2(d), 4(d) and 4(g)). The enlarged SSH chain in Fig. 4(a) has two features. One is the symmetry, which ensures that the effective hopping $t$ (illustrated in Fig. 4(h)) is identical between zeroth pseudo-LLs (see Fig. S8 for asymmetric ripples). The other is the varied amplitude, which excludes the possibility of connecting states localized in the ripples to form a snake-like pattern (Figs. S7 and S9).

To further explore effect of strain near the ripple head on the width of the zeroth pseudo-LLs as well as the effective hopping $t$, we carried out more calculation by reducing the strain near the ripple head with decreasing 25% of the amplitude of ripple head (other method is to modulate the buckling in SI Note 14), as shown in Fig. 4(e) of structure 2. The PMF of this structure, as shown in Fig. 4(f), shows lower intensity and the width of zero-mode increases from 180 meV to 250 meV with the corresponding hopping $t$ changing from 50 meV to 62.5 meV (LDOS in Fig. 4(g) and serpentine pattern in Fig. S10). This tendency can be inspired by the changes of nearest-neighbor hopping interactions near the ripple head. In the strained structures of the suspended graphene monolayer, localized states appear in both sub-lattice A and B of graphene (loss of sublattice polarization), the band width is determined by the effective hopping $t$ between adjacent zero modes, which is proportional to the nearest neighbor hopping amplitude of graphene. In fact, the effective hopping $t = 50{\sim}62.5$ meV is inside the energy range 5.2~120 meV with the minimum value corresponding to infinite $d_i$ and the maximum value corresponding to $d_i = d$ [40]. With increasing the strain, the band is flattened due to the reduced effective hopping strength.

In summary, by taking advantage of the nanoscale localization of the pseudo-LLs, we demonstrated the realization of the 1D SSH metallic states in graphene by using the strain-induced pseudo-LLs as the building block. Via STM and STS measurements, we directly image the 1D metallic states realized by periodic coupled zeroth pseudo-LLs along the 1D suspended graphene monolayer. Our result provides a new avenue to build custom-designed model systems with novel quantum phases by using nanoscale localized pseudo-LLs.

**Acknowledgements**

This work was supported by the National Natural Science Foundation of China (Grant Nos. 11974050, 11674029, 11921005, 11774269, 12047543) and National Key R and D Program of China (Grant No. 2017YFA0303301, 2018YFA0305800). L.H. also acknowledges support from the National Program for Support of Top-notch Young Professionals, support from "the Fundamental Research Funds for the Central Universities", and support from "Chang Jiang Scholars Program". Numerical calculations presented in this paper were performed on the supercomputing system in the Supercomputing Center of Wuhan University.

**Figures**

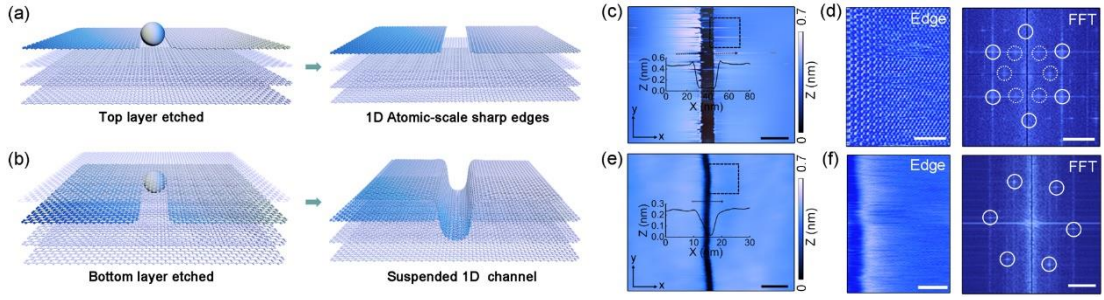

Fig. 1. (a) Schematic image of the top layer etched edges. (b) Schematic image of the bottom layer etched channel. (c) A STM topographical image of the sharp edges, $V_s = 1.4$ V, $I = 0.6$ nA. Inset: the height profile across the edge. Scale bar: 50 nm. (d) Left: the zoom-in atomic image of the square area in (c), $V_s = 0.4$ V, $I = 0.5$ nA. Scale bar: 2 nm. Right: The fast Fourier transform (FFT) image of one edge. The six outer spots are reciprocal lattices of graphene. The inner ($\sqrt{3} \times \sqrt{3}$) R30° pattern is induced by intervalley scattering from the edge. Scale bar: 3.5 nm$^{-1}$. (e) A Typical STM image of the bottom etched smooth channel, $V_s = -0.8$ V, $I = 0.2$ nA. Scale bar: 10 nm. (f) Left: the zoom-in image of the square area in (e), $V_s = 0.6$ V, $I = 0.1$ nA. Scale bar: 2 nm. Right: The FFT image of the top panel. The six spots are reciprocal lattice of graphene. Scale bar: 3.5 nm$^{-1}$.

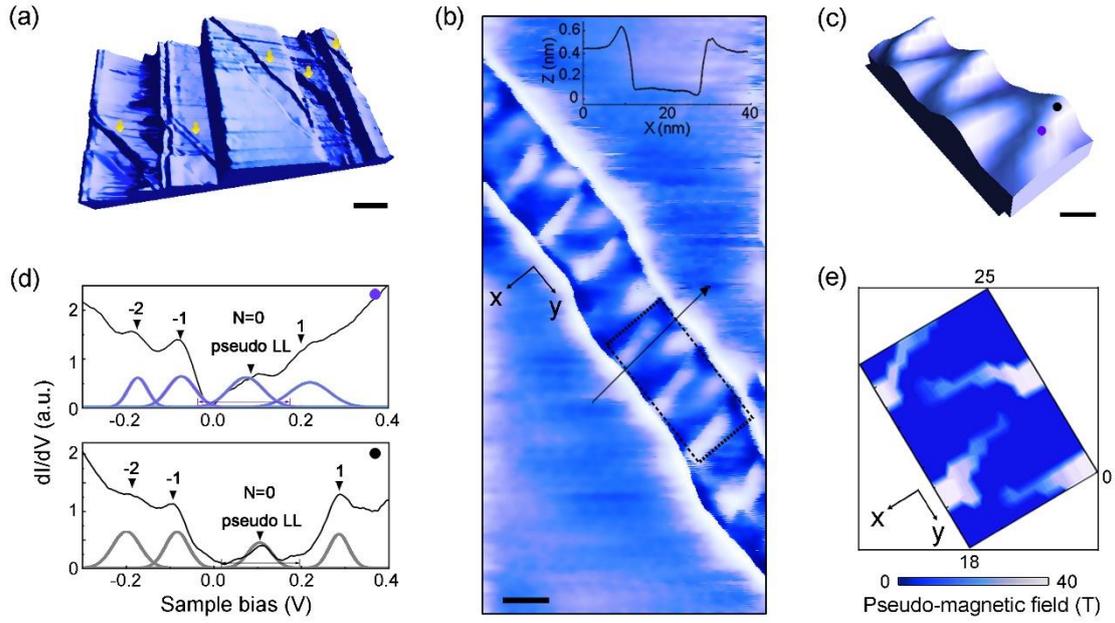

Fig. 2. (a) A 3D STM image of etched channels on continuous steps marked by yellow arrows, $V_s$ = -1.8 V, $I$ = 0.1 nA. Scale bar: 20 nm. (b) Two-dimensional projection of the longest channel in (a), $V_s$ = 0.05 V, $I$ = 0.015 nA. Scale bar: 10 nm. Inset: the height profile along the black arrow. (c) The zoom-in image of the black dashed square in (b). Scale bar: 3 nm. (d) The d$I$/d$V$ spectra taken around the center (top) and edge (bottom) of the ripple. The corresponding positions are marked by the purple and black dots in (c). (e) The distribution of PMFs along the 1D channel, showing the same period as the ripples.

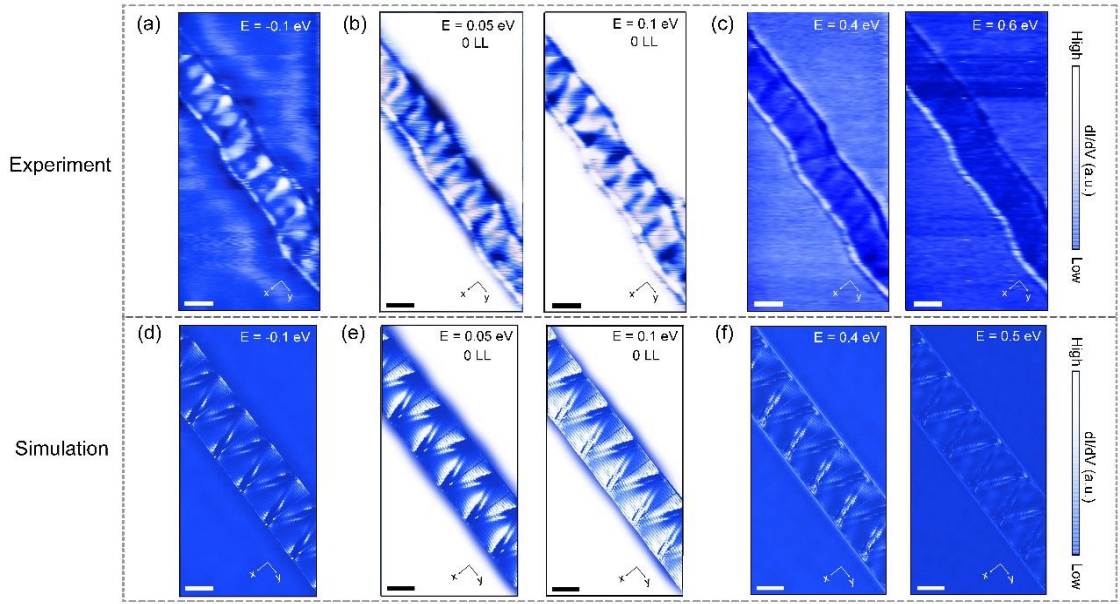

Fig. 3. The *dI/dV* maps measured at the energy of (a) -0.1 eV, (b) 0.05 eV and 0.1 eV (around zero pseudo-LL), (c) 0.4 eV and 0.6 eV, respectively. (d)-(f) Theoretical calculated *dI/dV* maps at the corresponding energies of -0.1 eV, 0.05 eV, 0.1 eV, 0.4 eV and 0.5 eV. The Dirac point is shifted to zero energy. Scale bar: 10 nm.

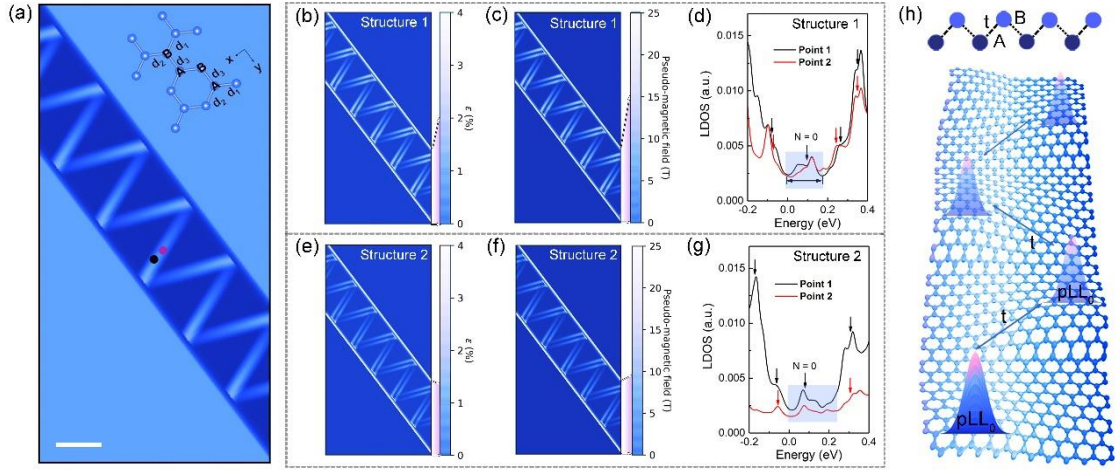

Fig. 4. (a) Schematic representation of the calculated graphene monolayer embedded with buckling and periodic ripples that have crests at an angle to the trench. The inset shows the sample has an armchair configuration along the channel. (b) and (e) Contour plots of the calculated average strain ε of the strained graphene structures 1 and 2, respectively. The pink areas denote the ranges of strain of two structures. (c) and (f) Calculated pseudo-magnetic field of structure 1 and 2, respectively. (d) and (g) Calculated local DOS of point 1 and point 2 along the ripple of structure 1 and 2, respectively. The corresponding positions are marked by the black and red dots in (a). (h) Top panel: Schematic representation of an effective SSH model to describe the strained graphene. A and B are two sublattices for a herringbone chain and *t* is the effective hopping between them. Bottom panel: Diagram of effective hopping *t* between two localized states (named pLL$_0$) in graphene.